\documentclass[aps,prb,twocolumn,preprintnumbers,amsmath,amssymb,bibnotes,showpacs,superscriptaddress,floatfix]{revtex4}
\usepackage{bbm}
\usepackage{mathrsfs}
\usepackage{amsmath}
\usepackage[dvips]{graphicx}
\usepackage{epsfig}
\usepackage{color}

\usepackage[utf8]{inputenc}

\newcommand {\ud}  {\mathrm{d}}
        
\renewcommand{\*}{\cdot}
\newcommand{\fr}[2]{\frac{#1}{#2}}

\newcommand{\bra}[1]{\langle #1 |}
\newcommand{\ket}[1]{| #1 \rangle}

\renewcommand{\vec}[1]{\boldsymbol{#1}}
\newcommand{\mc}{\mathcal}

\newcommand{\Id}{\mathbbm{1}}

\newcommand{\Tr}{\operatorname{Tr}}

\newcommand{\transp}{\intercal}

\newcommand{\id}{\mathbbm{1}}                       
\newcommand{\ham}{\hat{H}}                          
\newcommand{\obs}{\text{obs}}
\newcommand{\fit}{\text{fit}}

\newcommand{\aachen}{Institute for Theoretical Physics C, RWTH Aachen University, 52056 Aachen, Germany}
\newcommand{\potsdam}{Institute for Physics and Astronomy, Potsdam University, 14476 Potsdam-Golm, Germany}
\newcommand{\munich}{Physics Department and Arnold Sommerfeld Center for Theoretical Physics, Ludwig-Maximilians-Universit\"{a}t München, 80333 München, Germany}
\newcommand{\irvine}{Department of Physics and Astronomy, University of California, Irvine CA 92697, USA}

\begin{document}
\title{Spectral functions in one-dimensional quantum systems at $T>0$}

\author{Thomas Barthel}
\affiliation{\aachen}
\affiliation{\potsdam}
\author{Ulrich Schollw\"{o}ck}
\affiliation{\munich}
\author{Steven R. White}
\affiliation{\irvine}

\begin{abstract}
We present for the first time time-dependent density-matrix renormalization-group simulations ($t$-DMRG) at finite temperatures. It is demonstrated how a combination of finite-temperature $t$-DMRG and time-series prediction allows for an easy and very accurate calculation of spectral functions in one-dimensional quantum systems, irrespective of their statistics, for arbitrary temperatures. This is illustrated with spin structure factors of \emph{XX} and \emph{XXX} spin-$\frac{1}{2}$ chains. For the \emph{XX} model we can compare against an exact solution and for the \emph{XXX} model (Heisenberg antiferromagnet) against a Bethe Ansatz solution and quantum Monte Carlo data.
\end{abstract}

\date{January 15, 2009}

\maketitle

\section{Introduction}
Strongly correlated quantum systems continue to pose central challenges in theoretical condensed-matter physics. In the case of one-dimensional systems, we now have a full range of techniques to address static and dynamic ground-state properties. However, condensed-matter experiments typically work at finite temperatures that cannot be simply approximated by the ground-state physics, and the low-temperature physics of such systems is of high interest of its own. Due to the homogeneity of the systems under study in space and time, the experimental responses are best represented in momentum-frequency space, as, for example, by spin structure factors
\begin{equation}
S^{\mu\nu}(k,\omega) = \sum_{\ell}\, e^{ik\ell} \int \ud t \, e^{i\omega t} \langle\hat S_\ell^\mu (t)\hat S_0^\nu (0) \rangle ,
\end{equation}
where the average is taken with respect to some finite-$T$ density operator and $[\hat S^\mu_\ell,\hat S^\nu_{\ell'}]=i\delta_{\ell\ell'}\epsilon_{\mu\nu\gamma}\hat S^\gamma_\ell$. We will refer to functions of this form as spectral functions in the following; the proposed method does not depend on the specifics.
Theoretical approaches to calculate such spectral functions with high accuracy are very limited. On the other hand, experimental progress makes such calculations very timely: For example, due to small neutron flux, neutron-scattering techniques were in the past essentially limited to finding scattering maxima. Now, they have advanced to a degree that both the amplitude and the lineshape, which contain important information on the many-body physics involved, can be investigated with high accuracy (e.g.\ Refs.~\onlinecite{Stone2003-91} and~\onlinecite{Zaliznyak2004-93}).

For finite-temperature quantum Monte Carlo (QMC) calculations (e.g.\ positive-definite path integral \cite{Suzuki1977-58,Hirsch1982-26} or stochastic series expansion \cite{Sandvik1991-43} representation),  spectral functions have to be extracted by analytic continuation from imaginary-time results \cite{Jarrell1996-269}, which is ill-conditioned and numerically challenging.
In the context of DMRG \cite{White1992-11,White1993-10,Schollwoeck2005}, early approaches for finite-temperature spectral functions have built on transfer-matrix DMRG \cite{Shibata1997-66,Wang1997-56}, which gives thermodynamics for homogeneous infinite systems at high precision. In a first step, analytic continuation techniques as in QMC were employed \cite{Naef1999-60}. In a second development, transfer matrices were evolved in real-time to access autocorrelation functions \cite{Sirker2005-71,Sirker2006-73}; the time scales reachable limit resolution in frequency space.
While this approach would also be amenable to the prediction techniques used in the following to circumvent this issue, the determination of momentum-space correlators will be particularly easy with the $t$-DMRG methods\cite{Vidal2003-10,White2004,Daley2004} we use.
 In an unrelated development, a ground-state DMRG-like technique has been proposed that invokes a random-sampling and averaging procedure \cite{Sota2008-78}. Its feasibility for dynamical quantities has only been tested for very small systems; it is also limited to low temperatures.

In this paper, we present an application of $t$-DMRG at finite temperatures. It is shown how a combination of this technique and the linear prediction method \cite{Yule1927-226,Makhoul1975-63} allows one to obtain very accurate results for spectral functions with frequency and momentum dependence in the entire temperature range. While the demonstration focuses on a few particular cases, it will be obvious that the approach generalizes straightforwardly.

\section{Method}
\subsection{DMRG at finite temperature}
In recent years, DMRG \cite{White1992-11,White1993-10} has emerged as a key method for the ground-state physics of strongly correlated one-dimensional quantum systems \cite{Schollwoeck2005}. The time-evolution of pure states in one-dimensional strongly correlated spin, bosonic or fermionic models can now be simulated by $t$-DMRG \cite{Vidal2003-10,White2004,Daley2004} over times 10-100 times longer than the typical inverse energy scale of the problem (e.g.\ inverse hopping energy). It was quickly demonstrated \cite{Verstraete2004-6} that this method can be easily extended to the simulation of the static and dynamic behaviors at $T>0$ by using the purification \cite{Uhlmann1976,Uhlmann1986,Nielsen2000} of the density operator: Any density operator $\hat{\rho}$ of some physical system $\mc H$ can be encoded by a pure state of a combined physical and ancillary system, $|\rho\rangle\in\mc H\otimes\mc A$, such that the density matrix is retained by tracing out $\mc A$,
\begin{equation}
\hat{\rho} = {\rm tr}_{\mc A} |\rho\rangle\langle\rho|.	
\end{equation}
\begin{figure}[t]
\center{\epsfig{file=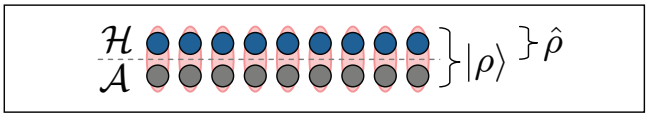,angle=0,width=\linewidth}}
\caption{(Color online.) The purification of a density operator $\hat\rho$ on a Hilbert space $\mc H$ is a pure state $\ket{\rho}$ on an enlarged Hilbert space $\mc H\otimes\mc A$ such that $\hat{\rho} = {\rm tr}_{\mc A} |\rho\rangle\langle\rho|$. The infinite temperature state $\hat \rho_0\propto \Id$ is obtained by preparing each pair of a physical site and the corresponding auxiliary site in a maximally entangled state.}
\label{fig:mps-purification}
\end{figure}
The ancillary system $\mc A$ can be taken as a copy of the physical state space $\mc H$; Fig.~\ref{fig:mps-purification}. In the case of interest, (unnormalized) thermal density operators $\hat{\rho}_\beta = e^{-\beta\ham}$, the corresponding purification can be constructed by an imaginary-time evolution starting from the
purification of the (trivial) infinite-$T$ ($\beta=0$) density operator $\hat{\rho}_0 = \id$. A possible purification for this $\beta=0$ ensemble is
\begin{equation}
  \ket{\rho_0}=\otimes_{\ell=1}^L \ket{\rho_{0,\ell}}
\quad\text{with}\quad
 \ket{\rho_{0,\ell}} =  \sum_{\sigma_\ell} \ket{\sigma_\ell}\otimes\ket{\sigma_\ell}',
\label{eq:maxentstate}
\end{equation}
where $\{\ket{\sigma_\ell}\}$ and  $\{\ket{\sigma_\ell}'\}$ denote the bases of the physical state space of site $\ell$ and its associated ancillary state space, respectively.
With this, we have $\hat{\rho}_0 = \Tr_{\mc A} \ket{\rho_0} \bra{\rho_0}$. Finite temperatures $\beta>0$ are  reached by imaginary-time evolution 
\begin{equation*}
\hat{\rho}_\beta = e^{-\beta\ham} = \Tr_{\mc A} \ket{\rho_\beta} \bra{\rho_\beta}
\quad\text{with}\quad
\ket{\rho_\beta}=e^{-\beta\ham/2} \ket{\rho_0}.
\end{equation*}
Proper normalization is restored by imposing $\langle\rho_\beta|\rho_\beta\rangle=1$. The mixed state $\hat \rho_\beta$ can then be evolved in time as $\ket{\rho_\beta (t)}=e^{-i\ham t} \ket{\rho_\beta (0)}$ and
$\hat{\rho}_\beta(t) = \Tr_{\mc A} \ket{\rho_\beta(t)} 
\bra{\rho_\beta(t)}$. 

As a product state, the initial $\beta=0$ purification $\ket{\rho_0}$ is uncorrelated, and hence, for the DMRG simulation, can be expressed exactly with block Hilbert spaces of dimension $m=1$. Imaginary-time evolution will introduce correlations, requiring one to increase $m$. For the evaluation of expectation values, both physical and ancillary degrees of freedom are traced over. As an example, take
\begin{eqnarray}
& &\Tr_{\mc H} (S^z_i(t) S^z_j(0) \hat{\rho}_\beta) = \Tr_{\mc H\otimes\mc A} (S^z_i(t) S^z_j(0) \ket{\rho_\beta}\bra{\rho_\beta}) \nonumber \\ & & = \bra{\rho_\beta} e^{i\ham t}S^z_i e^{-i\ham t} S^z_j \ket{\rho_\beta}.
\end{eqnarray}
While this approach has been found to yield thermodynamic quantities \cite{Feiguin2008-72} and static correlators \cite{Barthel2005} at $T>0$ to very high accuracy, real-time simulations are plagued by the same limitations as at $T=0$: the propagation of excitations through the system leads to entanglement growth in the purified state. As entanglement entropy is related roughly exponentially to DMRG resources, this strongly limits achievable simulation times, or inversely the $\omega$-resolution for spectral functions, as those are derived by Fourier transformation from the real-time data.
In order to circumvent this limitation at very low numerical cost, we adapt a linear prediction technique \cite{Yule1927-226,Makhoul1975-63} already successfully employed at $T=0$ in Ref.~\onlinecite{Pereira2008-100,White2008-77}.
\subsection{Linear prediction}\label{sec:linPred}
For a time series of complex data $x_0, x_1, \ldots, x_N$ at equidistant points in time $t_n=n\*\Delta t$ (and $t_\obs:= N \Delta t$) one makes a prediction of $x_{N+1}, x_{N+2}, \ldots$. In our case, $t_\obs$ is the (physical) time where the $t$-DMRG calculation was stopped -- usually because the computational cost to simulate further with a given accuracy has become too high.
For the data points beyond $t=t_\obs$, linear prediction makes the ansatz
\begin{equation}\label{eq:predictionAnsatz}
\tilde{x}_n = - \sum_{i=1}^p a_i x_{n-i} .
\end{equation}
The (predicted) value $\tilde{x}_n$ at time step $n$ is assumed to be a linear combination of $p$ previous values $\{x_{n-1},\dots,x_{n-p}\}$. Once the coefficients $a_i$ are determined from the known data, the ansatz is used to calculate (an approximation of) all $x_n$ with $n>N$.

The coefficients $a_i$ are determined by minimizing the least-square error in the predictions over a subinterval $(t_\obs-t_\fit,t_\obs]$ of the known data, i.e.\ we want to minimize
\begin{equation}\label{eq:predictionError}
 E\equiv\sum_{n\in \mc N_\fit } |\tilde x_n-x_n|^2/w_n
\end{equation}
where $\mc N_\fit$ is the fitting interval $\mc N_\fit=\{n| t_n \in (t_\obs-t_\fit,t_\obs]\}$, and $w_n$ is some weighting function; we choose $w_n\equiv 1$ for our simulations.
We found $t_\fit=t_\obs/2$ to be a robust choice. It is a compromise between choosing $t_\fit$ small enough to eliminate spurious short-time behavior from the true long-time behavior and choosing $t_\fit$ large to have a good statistics for the fit and allow for a large number $p$ of coefficients $a_i$ in the ansatz \eqref{eq:predictionAnsatz}.
Minimization of the error $E$, Eq.~\eqref{eq:predictionError}, with respect to the coefficients $a_i$ yields the system of linear equations
\begin{equation}
\label{eq:linPred:stationaryError}
R \vec{a} = -\vec{r},
\end{equation}
where $R$ and $\vec{r}$ are the autocorrelations
\begin{equation*}
R_{ji} = \sum_{n\in\mc N_\fit} x^*_{n-j} x_{n-i}/w_n,\quad
r_{j} = \sum_{n\in\mc N_\fit} x^*_{n-j} x_n/w_n.
\end{equation*}
Equation \eqref{eq:linPred:stationaryError} is solved by $\vec{a}=-R^{-1}\vec{r}$. For positive $w_n$, $R$ is a positive-definite matrix.

One may wonder why the extrapolation to infinite time is possible in this fashion. 
As demonstrated below, linear prediction generates a superposition of oscillating and exponentially decaying (or growing) terms, a type of time dependence that emerges naturally in many-body physics:
Green's functions of the typical form $G(k,\omega) = (\omega - \epsilon_k - \Sigma(k,\omega))^{-1}$ are in time-momentum representation dominated by the poles; e.g.\ for a single simple pole at $\omega=\omega_1 -i \eta_1$ with residue $c_1$, it will read
$G(k,t) = c_1 e^{-i\omega_1 t - \eta_1 t}$, and similarly it will be a superposition of such terms for more complicated pole structures. So the ansatz of the linear prediction is well suited for the typical properties of the response quantities we are interested in.
Note, however, that the method is not generally applicable for genuine nonequilibrium situations, as e.g.\ for the evolution of a local observable after a noninfinitesimal quench of system parameters. In such cases, typically, too many different frequencies can contribute to the signal, making the ansatz inappropriate.

To see the special form of time-series generated by the prediction, 
we introduce vectors $\vec{x}_n := [x_{n-1}, \ldots, x_{n-p}]^\transp$ such that \eqref{eq:predictionAnsatz} takes the form
\begin{equation}
\tilde{\vec{x}}_{n+1} = A \* \vec{x}_n,	
\end{equation}
 with 
\begin{equation}\label{eq:predictionMatrix}
A\equiv
\begin{bmatrix}
 -a_1&-a_2&-a_3&\cdots&-a_p\\
  1  & 0  & 0  &\cdots& 0  \\
  0  & 1  & 0  &\cdots& 0  \\
\vdots&\ddots&\ddots&\ddots&\vdots\\
  0   &\cdots& 0  & 1  & 0
\end{bmatrix}.
\end{equation}
Prediction therefore corresponds to applying powers of $A$ to the initial vector $\vec{x}_N$. A (right) eigenvector decomposition of $A$ with eigenvalues $\alpha_i$ leads to 
\begin{equation}
\tilde x_{N+m} = [A^m \* \vec{x}_N]_1=\sum_{i=1}^p c_i \alpha_i^m,
\end{equation}
where coefficients $c_i$ are determined from $\vec{x}_N$ and the eigenvectors of $A$. The eigenvalues $\alpha_i$ encode the physical resonance frequencies and dampings. The connection is given as
$\alpha_i = e^{i \omega_i \Delta t - \eta_i \Delta t}$.
Excluding exponentially growing terms as unphysical, all $\alpha_i$ should obey $| \alpha_i | \leq 1$. In numerical practice, eigenvalues $|\alpha_i|>1$ may occur, and various remedies exist, all based on the assumption that the corresponding $c_i$ are small. Common manipulations of occurring $|\alpha_i| > 1$ are $\alpha_i\to\alpha_i/|\alpha_i|$, $\alpha_i\to1/\alpha_i^*$, and $\alpha_i\to 0$. We tested all three and settled on the last option. When linear prediction is applicable, the choice should not matter (see discussion at the end of Sec.~\ref{sec:XXmodel}).

From the study of several examples, we found that the error of the linear prediction is roughly a function of $p\*\Delta t$, i.e.\ $p$ should be adapted to the choice of the time step $\Delta t$ in the DMRG time evolution. In the simulations, $p\*\Delta t=t_\fit/2$ was chosen as a compromise between a large $p$ to allow for a superposition of as many different oscillations as possible and a small $p$ to have good statistics for the fit of the coefficients $a_i$.

At $T=0$, critical one-dimensional systems exhibit power-law decays in their time-dependent correlators. The superposition of exponential decays is then taken to mimic these power-laws \cite{Pereira2008-100}.
At finite temperatures, time-dependent correlators $S(k,t)$ decay typically exponentially for large times (due to thermal broadening), making linear prediction especially well-suited for this situation.

\section{Applications}
\subsection{\emph{XX} model}\label{sec:XXmodel}
The Hamiltonian of the \emph{XX} model reads
\begin{equation}
\ham_{\text{XX}}=  \sum_i (\hat S^x_i \hat S^x_{i+1} + \hat S^y_i \hat S^y_{i+1}) + h\sum_i \hat S^z_i,
\end{equation}
where we choose the critical field $h=-1$.
We consider the observable
$S(\ell,t)=\fr{1}{2\pi} \langle [ \hat S^-_\ell(t),\hat S^+_0(0) ]\rangle_\beta$, which is the spectral function of the corresponding model of hardcore bosons. This allows for a detailed analysis of the method, as correlators for this model can be calculated numerically exactly by the evaluation of Pfaffian determinants \cite{Caianello1952,Stolze1995,Barthel2005}.
\begin{figure}[b]
\center{\epsfig{file=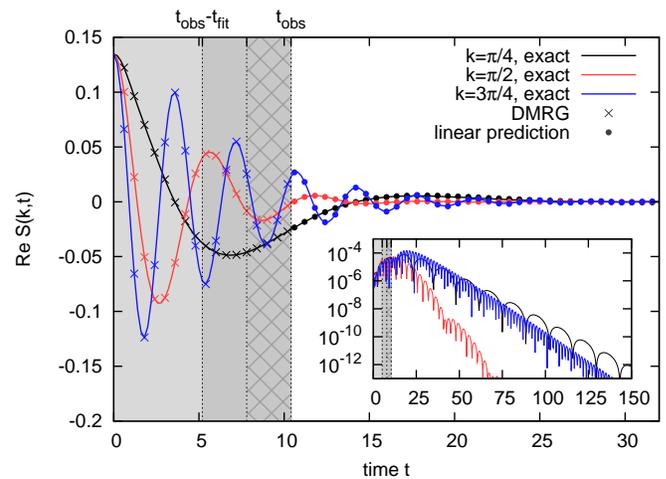,angle=0,width=\linewidth}}
\caption{(Color online.) \emph{XX} model at $\beta=10$: The exact $\text{Re\,} S(k,t)$ and the result of DMRG simulation followed by linear prediction are in excellent coincidence. The inset shows the long-time behavior of the deviation to the exact result. At $t=t_\obs$, the DMRG simulation was stopped (gray area). For the prediction, the deviation \eqref{eq:predictionError} of predicted to simulated values was minimized on the interval $(t_\obs-t_\fit,t_\obs]$ (dark gray) with respect to the coefficients $a_i$ in \eqref{eq:predictionAnsatz}. The hatched interval $(t_\obs-p\*\Delta t,t_\obs]$, determines the value of the first predicted value at $t=t_\obs+\Delta t$.}
\label{fig:realtimepredict}
\end{figure}

The DMRG simulation employed a fourth-order Trotter-Suzuki decomposition and a time step $\Delta t=0.2$. Deviations $||\ket{\rho_{\Delta t}^\text{exact}}-\ket{\rho_{\Delta t}^\text{DMRG}}||/\Delta t$ between exactly time-evolved state and its DMRG approximation were bounded from above by $0.005$ per time unit. We used a lattice of $L=100$ sites 
\footnote{\label{foot:finitesize} The finite temperature causes a finite correlation length in the $t=0$ state. During time evolution, correlations are built up further only inside causal cones.  Hence for small enough $\beta$ and $t$, finite-size effects are negligible if measurements are restricted to the middle of the system.}.
To demonstrate the potential of the prediction method, the calculations was stopped at times $t_\obs=10.4$ for $\beta=10$ and $t_\obs=26.6$ for $\beta=50$ (which can be reached with very moderate computational resources corresponding to DMRG block Hilbert spaces of dimension $m \lesssim 200$).
For the linear prediction \eqref{eq:predictionAnsatz}, we used $p=14$ for $\beta=10$ and $p=34$ for $\beta=50$. The fitting interval was taken from half the maximal computation time onward ($t_\fit=t_\obs/2$ and $p\*\Delta t=t_\obs/4$).
\begin{figure}[t]
\center{\epsfig{file=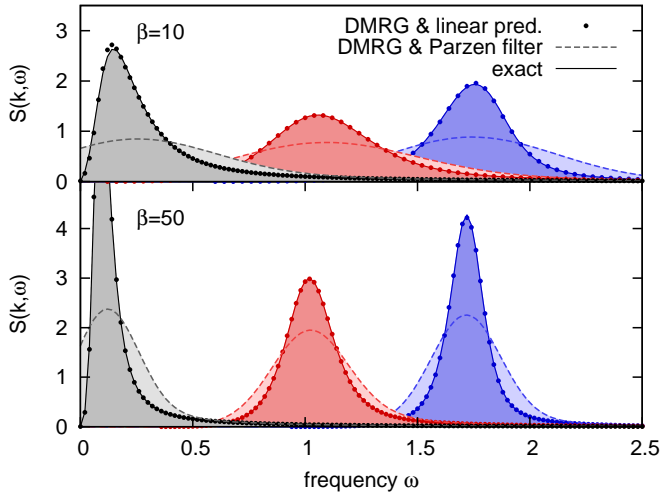,angle=0,width=\linewidth}}
\caption{(Color online.) Lineshapes for the \emph{XX} model at various values of the quasimomentum $k$ (from left to right: $k=\pi/8$, $\pi/3$, and $3\pi/4$) at $\beta=10$ and $\beta=50$ : exact (solid lines) and predicted (dots) lineshapes are in excellent agreement. For comparison, Parzen-filtered lineshapes (dashed lines) obtained from (raw) DMRG data are shown.}
\label{fig:lineshapes}
\end{figure}

In Fig.~\ref{fig:realtimepredict} it can be seen that very accurate predictions are possible far into the future. As Fig.~\ref{fig:lineshapes} shows, for both temperatures, lineshapes can be calculated very accurately upon Fourier transformation of the predicted data, allowing for precise experimental analysis. In contrast, even upon windowing (here done by a Parzen filter), the raw data provide completely wrong line maxima and widths. 
Almost the whole spectral function is reproduced very exactly. The sole exception occurs for values of $k$ very close to zero and $\pi$: here, the group velocity vanishes, $\ud\omega/\ud k\rightarrow 0$, leading to a very slow decay of correlators which is not captured perfectly for the chosen temperatures and maximum simulation times $t_\obs$.

In Sec.~\ref{sec:linPred}, we gave already arguments for the parameter choice $t_\fit=t_\obs/2$ and $p\*\Delta t=t_\fit/2$ for the prediction method. There is one further parameter involved: For the determination of prediction coefficients $a_i$ in \eqref{eq:predictionAnsatz}, the autocorrelation matrix $R$ has to be inverted, \eqref{eq:linPred:stationaryError}, and requires some regularization.
To this purpose, one may either add a small regularization constant before the inversion $R^{-1}\to(R+\varepsilon \Id)^{-1}$, or project out the eigenspaces with eigenvalues below $\varepsilon$ i.e.\ $R^{-1}\to (P_\varepsilon R P_\varepsilon)^{-1}$ (pseudo inverse).

In cases where the linear prediction method is not appropriate, or the $t_\obs$ achievable with $t$-DMRG is too small, the results of the prediction are generally sensitive to variation of the parameters $t_\fit$, $p$, and $\varepsilon$. In this respect, the pseudo inverse approach $R^{-1}\to (P_\varepsilon R P_\varepsilon)^{-1}$ is favorable, as it produces stronger variations in the result for small changes in $\varepsilon$, if the linear prediction is not applicable. If the method is applicable, the regularization parameter $\varepsilon$ can be varied over several orders of magnitude without effect, but should of course be chosen as small as possible ($10^{-7}$--$10^{-6}$ in our calculations).
An efficient procedure to fix $\varepsilon$ is to examine plots of the eigenvalues $\alpha_i(k)$ of \eqref{eq:predictionMatrix} as functions of the quasimomentum $k$ for several $\varepsilon$. For too small $\varepsilon$, noisy scatter and many abrupt jumps appear, resulting in completely wrong lineshapes for some $k$-values; $\varepsilon$ is sufficiently big when $\alpha_i(k)$ shows smooth ``band structures" with only a few abrupt jumps. In cases where this is only possible with large $\varepsilon$, the prediction method is not applicable.

\subsection{Isotropic Heisenberg spin-$\fr{1}{2}$ antiferromagnet}
For the Heisenberg antiferromagnet (XXX model)
\begin{equation}
\ham_{\text{HAFM}} = \sum_i \hat{\vec{S}}_i \* \hat{\vec{S}}_{i+1}
\end{equation}
we have calculated the longitudinal spin structure factor (spectral function) $S^{zz}(\ell,t)=\langle \hat S^z_\ell(t)\hat S^z_0(0)\rangle_\beta$. This case is interesting and challenging as already at $T=0$ a whole continuum of excitations contributes to the spin structure factor as opposed to the simple dispersion of the \emph{XX} model with a single peak in $S(k,\omega)$ for each momentum.
As concluded from ($T=0$) Bethe ansatz calculations, the dominant contributions stem from a two-spinon continuum bounded from below and above by \cite{Cloizeaux1962-128,Yamada1969-41}
\begin{equation}
\label{eq:HAFMspectrumBounds}
 \varepsilon_L(k)=\fr{\pi}{2}|\sin{k}|
\quad\text{and}\quad
 \varepsilon_U(k)=\pi|\sin{\fr{k}{2}}|.
\end{equation}
The exact contribution of the two-spinon continuum was derived in Ref.~\onlinecite{Bougourzi1996-54,Karbach1997-55} and the contribution of the four-spinon continuum computed in Ref.~\onlinecite{Caux2006-12} (see also Ref.~\onlinecite{Caux2005-09}).

The $t$-DMRG calculation was done for a lattice of 128 sites\footnotemark[\value{footnote}], with a fourth-order Trotter-Suzuki decomposition and time steps $\Delta \beta=\Delta t=0.125$. The truncation error $\chi^2=\sum_{i>m}\lambda_i^2$ ($\chi$ is the 2-norm of the discarded Schmidt coefficients $\lambda_i$) was chosen as $\chi^2\leq 10^{-12}$ during the cooling and $\chi^2\leq 10^{-10}$ during the real-time evolution, resulting in a maximum DMRG block Hilbert-space dimension of $m\lesssim 1200$ for the maximum simulation times $t_\obs=6.25,7.25,10.5$ at $\beta=1,4,16$, respectively.
For the linear prediction \eqref{eq:predictionAnsatz}, we used correspondingly $p=13,15,21$ ($p\*\Delta t=t_\obs/4$) and $t_\fit=t_\obs/2$.

\begin{figure}[t]
\epsfig{file=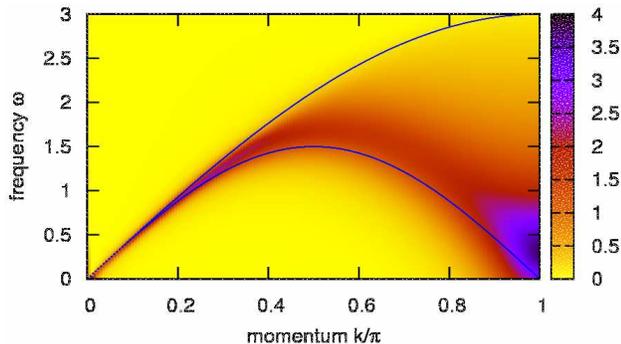,angle=0,width=0.95\linewidth}
\caption{(Color online.) The longitudinal spin structure factor  $S^{zz}(k,\omega)$ for the isotropic Heisenberg antiferromagnet at $\beta=4$ as obtained from DMRG time evolution followed by linear prediction with $t_\obs=7.25$, $t_\fit=3.5$, and $p=15$. Blue lines mark the bounds \eqref{eq:HAFMspectrumBounds} on the ($T=0$) two-spinon spectrum.}
\label{fig:HAFM}
\end{figure}

\begin{figure}[t]
\epsfig{file=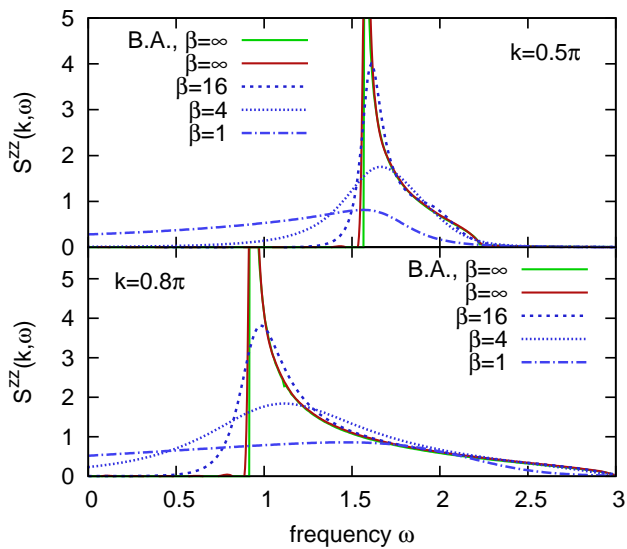,angle=0,width=0.95\linewidth}
\caption{(Color online.) Line shapes of the longitudinal spin structure factor  $S^{zz}(k,\omega)$ for the isotropic Heisenberg antiferromagnet as obtained from DMRG and linear prediction at finite $T$, compared to the $T=0$ Bethe ansatz result. The Bethe ansatz data (``B.A.'', provided by J.-S. Caux \cite{Caux2006-12}) comprises the two- and four-spinon contributions, accounting for about 98\% of the total weight.}
\label{fig:HAFM-lineShapes}
\end{figure}

The result of DMRG simulation, combined with linear prediction is displayed in Figs.~\ref{fig:HAFM} and \ref{fig:HAFM-lineShapes}. The structure factor converges for $\beta\to\infty$ to the $T=0$ Bethe ansatz results from Ref.~\onlinecite{Caux2006-12}.


\begin{figure}[t]
\epsfig{file=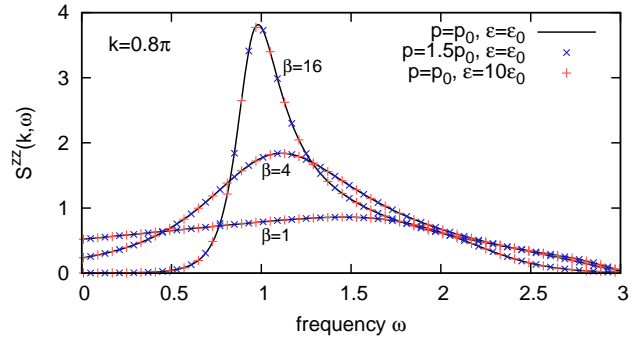,angle=0,width=0.95\linewidth}
\caption{(Color online.) Testing the robustness of the linear prediction method at the example of  $S^{zz}(k,\omega)$. The black curves ($p=p_0$ and $\varepsilon=\varepsilon_0$) correspond to the same DMRG data as in Fig.~\ref{fig:HAFM-lineShapes} ($\Delta \beta=\Delta t=0.125$; for $\beta=1,4,16$, we used $p_0=13,15,21$ and $\varepsilon_0=10^{-7},10^{-7},10^{-6}$, respectively). The blue tilted crosses correspond to simulations where the length $p\*\Delta t$ of the prediction interval was increased by a factor of 1.5. The red crosses correspond to calculations where the regularization parameter $\varepsilon$ for the inversion of the autocorrelation matrix $R$ was increased by a factor of 10.}
\label{fig:HAFM-lineShapes-robustness}
\end{figure}

Fig.~\ref{fig:HAFM-lineShapes-robustness} illustrates robustness of the linear prediction against variation of the corresponding parameters. Changing the regularization parameter by a factor of 10 or the length $p\*\Delta t$ of the prediction interval by a factor 1.5 has no visible effect.
\begin{figure}[t]
\epsfig{file=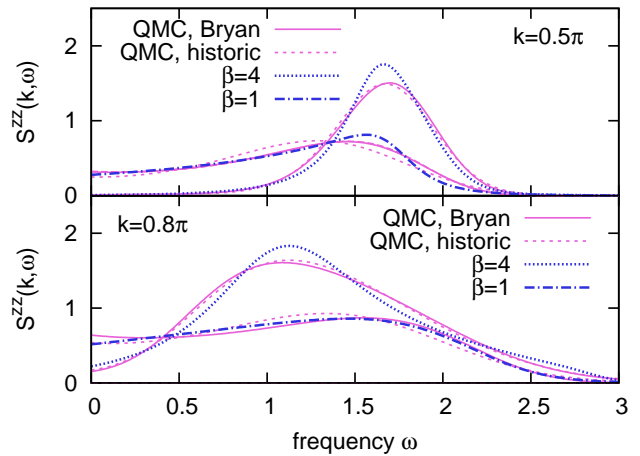,angle=0,width=0.95\linewidth}
\caption{(Color online.) Comparison of the longitudinal spin structure factor  $S^{zz}(k,\omega)$ for the isotropic Heisenberg antiferromagnet as obtained from DMRG and linear prediction on the one hand, and QMC and the maximum entropy method on the other hand (provided by S.~Grossjohann \cite{Grossjohann2009}). The maximum deviation is about 20\%. Labels ``Bryan'' and ``historic'' refer to the maximum entropy methods of Refs.~\onlinecite{Skilling1984-211} and Ref.~\onlinecite{Gubernatis1991-44}, respectively.}
\label{fig:HAFM-lineShapes-QMC}
\end{figure}

\begin{figure}[t]
\epsfig{file=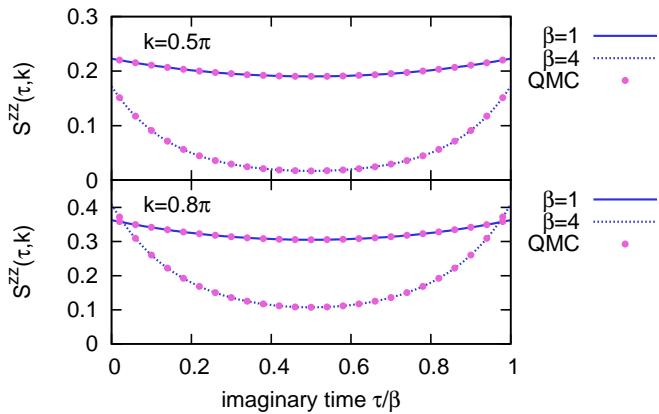,angle=0,width=1\linewidth}
\caption{(Color online.) Comparison of the longitudinal spin structure factor  $S^{zz}(k,\tau)$ on the imaginary-time axis. DMRG with linear prediction (lines) and QMC \cite{Grossjohann2009} (dots) agree very well (maximum discrepancy of $\sim 5\*10^{-4}$). It is to be concluded that the deviations in $S^{zz}(k,\omega)$ stem from the ill-conditioned analytic continuation which has to be carried out for the QMC data and amplifies small discrepancies.}
\label{fig:HAFM-lineShapes-QMC-imag}
\end{figure}

There is no Bethe ansatz result available for $T>0$. We hence compare in Figs.~\ref{fig:HAFM-lineShapes-QMC} and \ref{fig:HAFM-lineShapes-QMC-imag} to recent quantum Monte Carlo (QMC) data\cite{Grossjohann2009}. QMC simulations yield the spin structure factor $S^{\mu\nu}(k,\tau)$ on the imaginary-time axis. This observable can be determined to very high precision. The actual quantity of interest is however $S^{\mu\nu}(k,\omega)$, where
\begin{equation}
\label{eq:imaginary-time-S}
S^{\mu\nu}(k,\tau)=\fr{1}{2\pi}\int_{-\infty}^\infty\ud\omega e^{-\omega\tau}S^{\mu\nu}(k,\omega).
\end{equation}
To obtain $S^{\mu\nu}(k,\omega)$ from $S^{\mu\nu}(k,\tau)$, one can employ several variants of the maximum entropy method (see e.g.\ Ref.~\onlinecite{Jarrell1996-269} for a review).
The longitudinal spin structure factor $S^{zz}(k,\omega)$, Fig.~\ref{fig:HAFM-lineShapes-QMC}, shows a maximum deviation of about 20\%. If one compares however on the imaginary-time axis, $S^{zz}(k,\tau)$ given in Fig.~\ref{fig:HAFM-lineShapes-QMC-imag}, QMC and DMRG with linear prediction (the DMRG data was transformed according to \eqref{eq:imaginary-time-S}) agree with a maximum deviation of $\sim 5\*10^{-4}$. The discrepancy of the corresponding data for $S^{zz}(k,\omega)$, is hence to be attributed to the fact that the maximum entropy method, employed to transform the QMC data to frequency space, is ill-conditioned. This is due to the exponential in \eqref{eq:imaginary-time-S}. Small errors of $S^{zz}(k,\tau)$ get blown up by the numerical analytic continuation.

\section{Discussion}
We have demonstrated how a combination of finite-temperature $t$-DMRG and linear prediction allows for a very accurate calculation of spectral functions, at finite temperatures. The method is for one-dimensional systems favorable to the usual QMC approach: As $t$-DMRG has no sign problem, fermionic systems are also directly accessible. Also, no analytic continuation of correlators as in QMC calculations is necessary. As opposed to some other approaches, the proposed method works over the entire temperature regime. An attractive feature of the method is provided by the increasing availability of $t$-DMRG codes -- which need only minor modification to simulate at finite temperatures -- and the simplicity of the numerically inexpensive prediction method.

\acknowledgments
This work was supported by the DFG and the Studienstiftung des Deutschen Volkes (T.~B.) as well as the NSF under grant DMR-0605444 (S.~R.~W.). We are grateful to J.-S.\ Caux for providing the Bethe ansatz data for Fig.~\ref{fig:HAFM-lineShapes} and to S.~Grossjohann for the QMC data for Figs.~\ref{fig:HAFM-lineShapes-QMC} and \ref{fig:HAFM-lineShapes-QMC-imag}.

\bibliographystyle{prsty} 

\end{document}